# ANALYSIS OF AUTO ZEROING AMPLIFIER

S.Anand, S.Porkodi
Department of ECE
Mepco Schlenk Engineering College,
Sivakasi-626005, India
sanand@mepcoeng.ac.in

**ABSTRACT:**

This paper presents a new auto zeroing technique that combines very high-speed operation, low power consumption, and low input switching interferences. This technique has been applied to the design of CMOS flash Analog-to-Digital converter for Ultra-Wide Band applications. This project is dealt with the design of an auto zeroing amplifier with three stages. It has to be operating on 1.2 power supply voltage and the input voltage is limited to 400 mv in 130 nm CMOS technology. The CMOS auto zeroing amplifier stages are simulated in the Advanced Design System (ADS) Tool.

## 1. INTRODUCTION

A wide variety of electronic applications deals with the conditioning of small input signals. These systems require very low offset voltage [1]. The amplifiers with by far the lowest offset are the Auto Zero Amplifiers (AZA). These amplifiers achieve high DC precision through a continuously running calibration mechanism. Today's Auto zeroing amplifiers neither in form nor in the application from standard operational amplifiers. An auto zeroing amplifier is a self-calibrating amplifier in the beginning stage of the self-calibrating amplifier is called chopper amplifiers. It provides the very low offset voltage but designing of these amplifiers will lead to more complexity. To overcome these complexities AZA can be used. This AZA is used in the high precision circuits with wide bandwidth and low output noise [2], [3], [4]. The use of AZA as a,

- ➢ Signal amplifier
- ➢ Calibrating amplifier in dc and
- ➢ A wideband amplifier in ac

The auto zeroing technique is widely applied to reduce the comparator's offset error. it provides the high open-loop gain for preamplifiers and also it is used to reduce the front end amplifiers. The AZA design is not available in the integrated circuits but requires multiple amplifier integrated circuits instead. It performs the offset correction by a 'sample and hold' method. The proposed scheme does not have capacitors at the ADC input, reducing the input loading, and thus the power consumption The AZA circuit design has reduced the amount of supply voltage, allowing for very low operation in high gain, high precision applications. The integration of the external storage capacitors in combination with the performance of the AZA makes AZA is used to use as standard CMOS op-amps.

## 2. AUTO ZEROING TECHNIQUE

The operating principle of the conventional auto zeroing technique is described as the two non-overlapped control signals $\phi_1$ and $\phi_2$ are needed to avoid the connection between the input signal and the reference voltage. Through the alternate operation of the signals $\phi_1$, and $\phi_2$, the offset error of the comparator is periodically sensed, stored, and subtracted from the input signal.

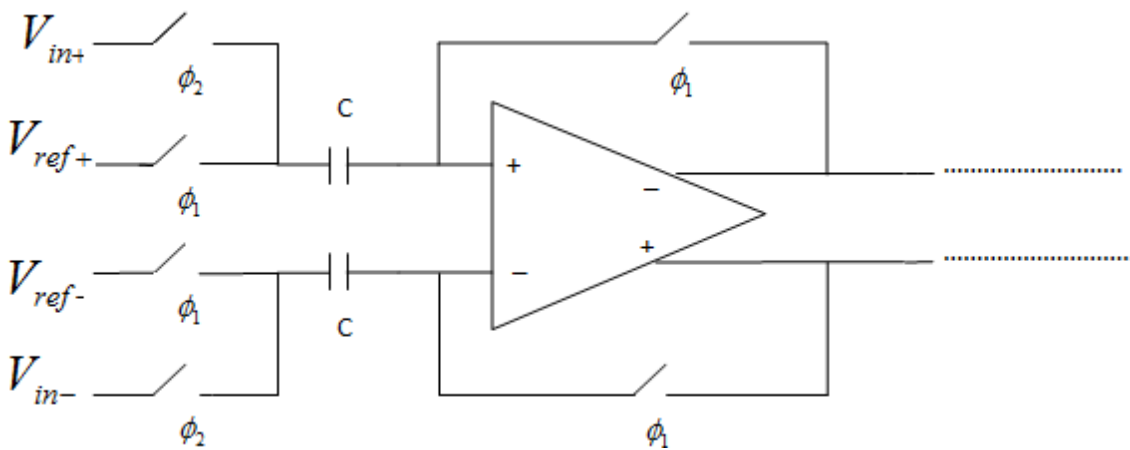

**Fig.1 The input stage of the conventional amplifier with auto zeroing technique**

The Fig.1 represents the Input stage of the conventional amplifier with auto zeroing technique. In this amplifier during the first clock phase ($\phi_1$), the difference between the reference voltage and the amplifier offset is stored in capacitor C. In the second clock phase ($\phi_2$) this value is subtracted from the input signal and applied to the open-loop amplifier.

This conventional auto zeroing technique has important drawbacks as [5], [6], [7], [8],

- ➢ Input switching interference
- ➢ Existence of two nonoverlapping clock phases.

## 2.1 NEW AUTO ZEROING TECHNIQUE

The errors are produced by the conventional auto zeroing technique are overcome by a new auto zeroing technique as Fig.2.

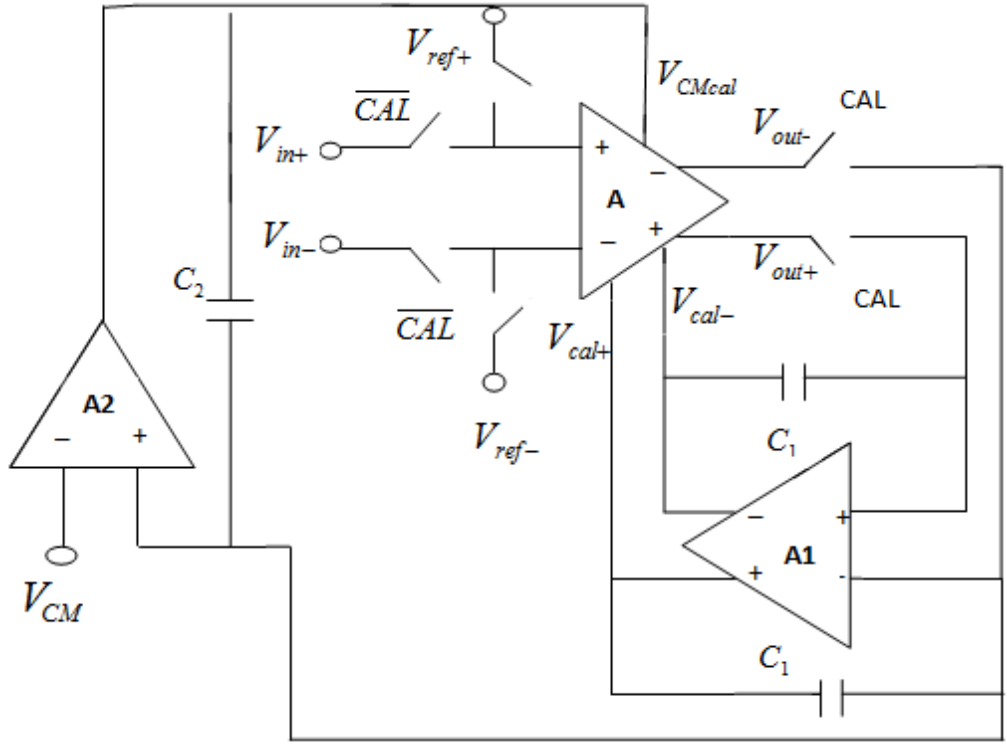

**Fig.2 Amplification stage with new Auto Zeroing technique**

This auto zeroing technique consists of three amplifier stages that are described by amplifiers A (main amplifier), A1 and A2. It operates in two different modes

- ➢ Calibration mode (CAL = 1)
- ➢ Normal operation mode (CAL = 0)

## 2.2 CALIBRATION MODE

In this mode, a control signal CAL is at a high level and the differential reference voltage ($V_{ref+}$ (-) $V_{ref-}$) obtained from a resistor ladder, is applied to the inputs of the main amplifier. It can be noted that if $V_{ref}$ is too large, the linearity of the amplifier can be compromised.

2.3 NORMAL OPERATION MODE

Once the calibration phase has finished, the CAL signal goes down, so that the differential signal coming from the previous stage (or the input signal, for the case of the first stage) is applied to the inputs of the main amplifier. During this phase, capacitors C1 and C2 are disconnected from the amplifier outputs and their respective feedback loops remain open; therefore, it is possible the operation of the amplifier at very high frequency. The values of C1 and C2 should be large enough to maintain the calibration for a long time.

## 3. DESIGN OF AUTO ZERO AMPLIFICATION STAGES

To design of AZA stage consists of the three amplifiers. In the designing of the three amplifiers gain and bandwidth are calculated.

### 3.1 AMPLIFIER A2

Operational amplifiers are an integral part of many analog and mixed signal systems. Amplifier A2 can be designed to have a low offset and a high dc gain using conventional two-stage amplifiers. Generally, Two-stage OP-AMP mainly consists of a cascade of Voltage to Current and Current to voltage stages. The first stage consists of a differential amplifier converting the differential input voltage to differential currents. These differential currents are applied to a current mirror load recovering the differential voltage. The second stage consists of common source MOSFET converting the second stage input voltage to current. This transistor is loaded by a current sink load, which converts the current to the voltage at the output.

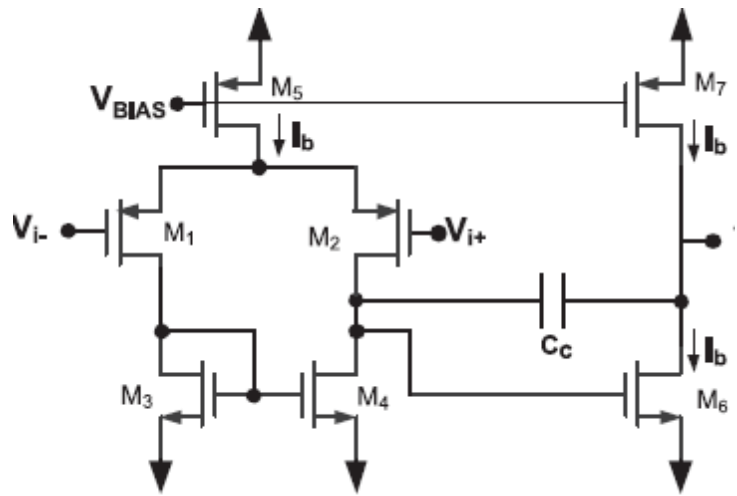

**Fig.3 The architecture of the A2 amplifier**

Fig.3 denotes the CMOS Architecture for the A1 amplifier and consists of two stages.

- ➢ Differential amplifier
- ➢ Common source amplifier

3.1.1 Differential Amplifier: The differential amplifier is a basic building block of an op-amp. The function of a differential amplifier is to amplify the difference between two input signals. In the A2 amplifier, the MOSFET m1-m2 and m3-m4 pair will act as PMOS input pair differential amplifier. Generally, the differential amplifier is divided into 4 categories. There are,

- ➢ Dual input, balanced output differential amplifier.
- ➢ Dual input, unbalanced output differential amplifier.
- ➢ Single input balanced output differential amplifier.
- ➢ Single input unbalanced output differential amplifier

A dual input, unbalanced output differential amplifier type differential amplifier is used in the A2 amplifier. But, it produces the low gain. A PMOS m5 is used to drive the input bias voltage to the circuit.

3.1.2 Common Source Amplifier: A common source amplifier is a second stage of the two-stage operational amplifier. It can be connected with the first stage amplifier by the miller compensation capacitance $C_c$ for the simplest frequency compensation. The gain of the second stage will multiply with the first stage gain to produce a high gain of the A2 amplifier. A MOSFET pair m6-m7 will act as a common source amplifier stage. The value of $C_c$ is an important factor when determining noise and power. Power consumption $C_c$ can be reduced but at the expense of noise performance.

3.2 AMPLIFIER A1

Amplifier A1 is the low frequency, high gain amplifier circuit in the auto zeroing technique. It is also a two-stage based op-amp. But it consists of three stages, so it is called multi stage amplifier. With two stage architecture additionally two MOSFET and a coupling capacitor is added Fig.4.

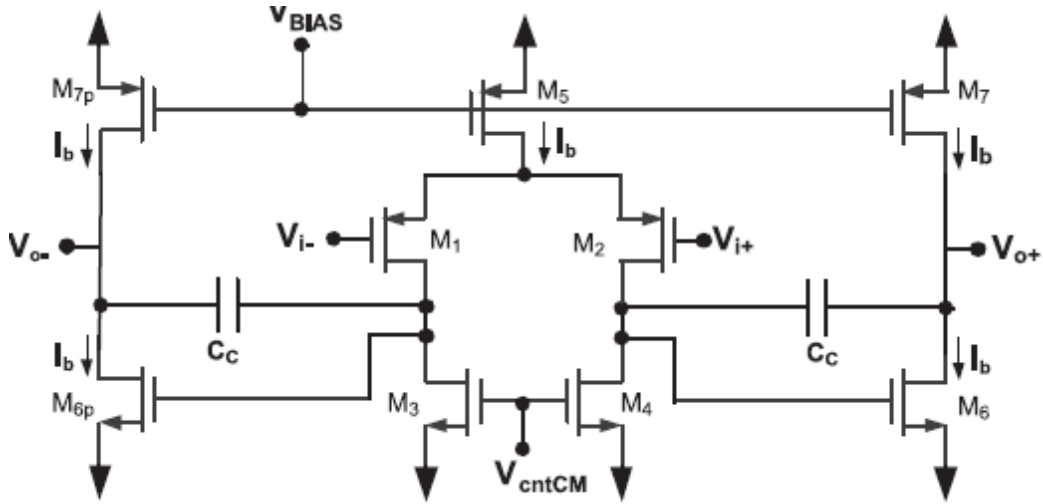

**Fig.4 The architecture of the A1 amplifier**

      A control common voltage is applied between the M3-M4 pair. A bias voltage is applied to the M5-M7-M7 transistor. The coupling capacitors are added on both sides of the design for frequency compensation. It is also a high gain low frequency amplifier.

### 3.3 MAIN AMPLIFIER (A)

      The main amplifier of the auto zeroing technique is denoted as amplifier A.

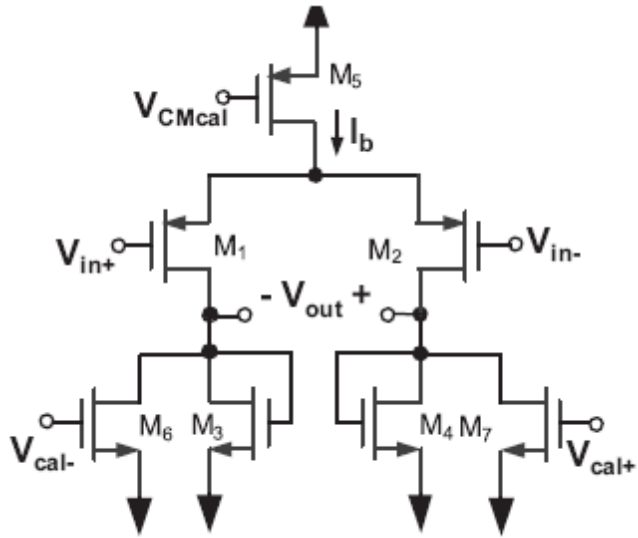

**Fig.5 The architecture of the main amplifier**

      Fig.5. represents the main amplifier architecture. The (M1, M2) are the differential pair with the diode-connected load (M3, M4). Transistors M6 and M7 are used to compensate for its offset error $V_{off}$. The common-mode voltage is controlled by means of the gate terminal of M5. A calibration voltage is applied to the gate voltage of the M6 and M7 transistor. The positive and

negative terminal of the calibration voltage is the output voltages from the A1 amplifier. By properly designing this voltage value the offset error of the main amplifier will be reduced. The main amplifier is a high frequency amplifier and it has a low gain compare with the two amplifier stages.

## 4. SIMULATION AND RESULTS

By applying a proper supply voltage, input voltage, and width of the transistors the gain of the three amplifiers is calculated. The simulation was done by ADS tool and CMOS technology used is 130 nm CMOS technology. A MOSFET model used for this project is BSIM 4 model. The gain waveforms of the three amplification stages are described below. A2 and A1 amplifiers are operating at a low frequency and the gain value is high. These 2 amplifiers are used to reduce the offset of the main amplifier. The main amplifier can operate on high frequency and the gain value is low compare to the other two amplifiers as in Fig .6, Fig.7, Fig.8.

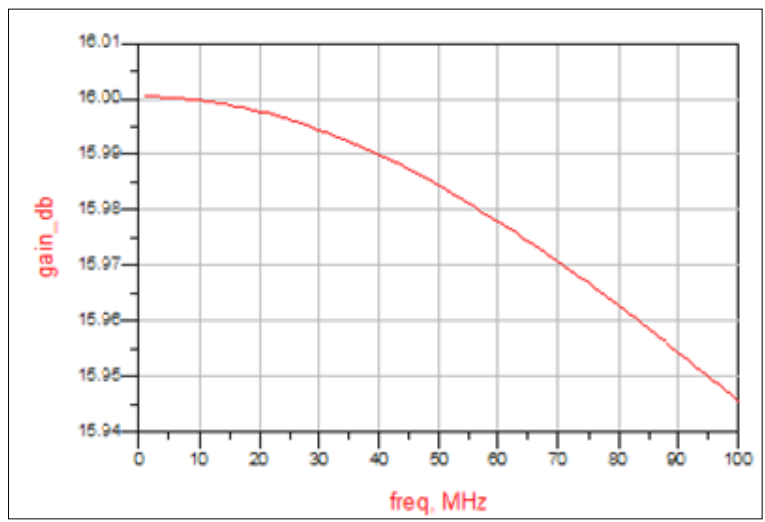

**Fig.6 The gain waveform for A1 Amplifier**

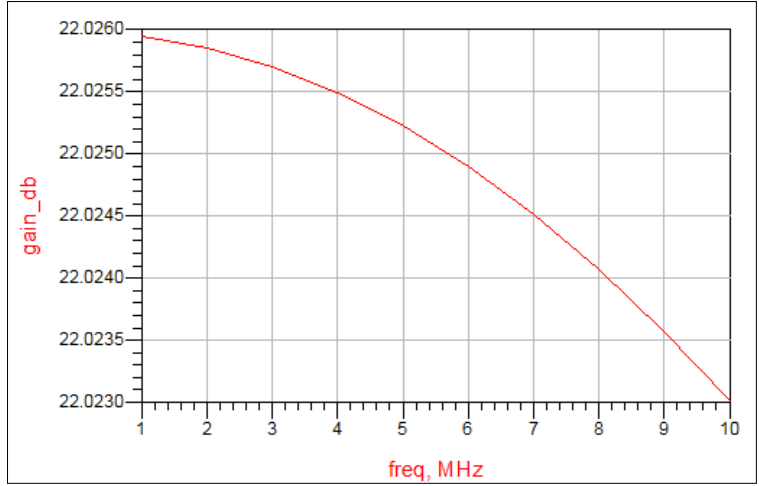

**Fig.7 The gain waveform for A1 Amplifier**

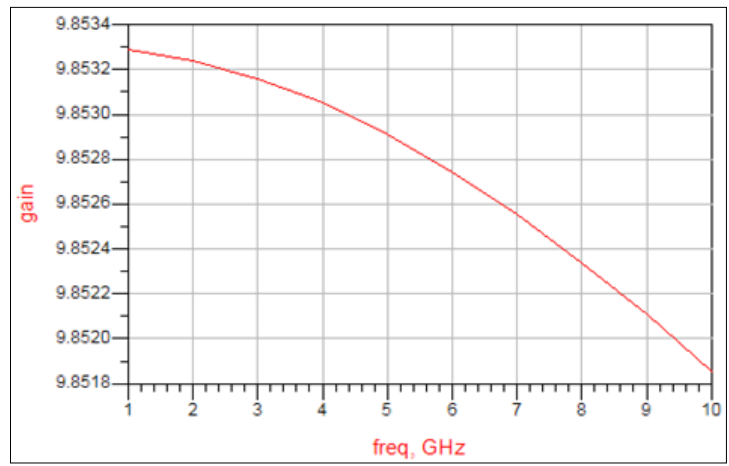

**Fig.8 The gain waveform for A1 Amplifier**

By properly designing the three amplification stages circuits the gain was calculated in the frequency range.

## 5. CONCLUSION

A new auto zeroing technique that combines very high-speed operation, low power consumption, and low input switching interferences has been analyzed. This technique can be applied to the design of CMOS flash Analog-to-Digital converter for Ultra-Wide Band applications. This analysis dealt with the design of an AZA with three stages and operated on 1.2 power supply voltage and the input voltage is limited to 400 mv in 130 nm CMOS technology. The gain waveforms of A1 and A2 are obtained. The CMOS auto zeroing amplifier stages are simulated in the ADS Tool.

**REFERENCES**

[1]. Iwamoto, Motomitsu. "Auto-zero amplifier and feedback amplifier circuit using the auto-zero amplifier." U.S. Patent No. 8,922,276. 30 Dec. 2014.

[2]. Kusuda, Yoshinori. "A 5.9 nV/√ Hz chopper operational amplifier with 0.78 μV maximum offset and 28.3 nV/° C offset drift." 2011 IEEE International Solid-State Circuits Conference. IEEE, 2011.

[3]. Anand, S., and Keetha Manjari RK. "FPGA implementation of artificial neural network for forest fire detection in wireless sensor network." 2017 2nd International Conference on Computing and Communications Technologies (ICCCT). IEEE, 2017

[4]. Kusuda, Yoshinori. "A 60 V Auto-Zero and Chopper Operational Amplifier With 800 kHz Interleaved Clocks and Input Bias Current Trimming." IEEE Journal of Solid-State Circuits 50.12 (2015): 2804-2813.

[5]. McCartney, Damien. "Auto-zero switched-capacitor integrator." U.S. Patent No. 5,479,130. 26 Dec. 1995.

[6]. Raghuveer, V., Karthi Balasubramanian, and Singamala Sudhakar. "A 2μV low offset, 130 dB high gain continuous auto zero operational amplifier." 2017 International Conference on Communication and Signal Processing (ICCSP). IEEE, 2017.

[7]. Huijsing, Johan. "Low-Noise and Low-Offset Operational and Instrumentation Amplifiers." Operational Amplifiers. Springer, Cham, 2017. 351-413.

[8]. Premkumar, R., and S. Anand. "Secured and compound 3-D chaos image encryption using hybrid mutation and crossover operator." Multimedia Tools and Applications 78.8 (2019): 9577-9593.